\documentstyle[amssymb,aps,epsfig,float,twocolumn]{revtex}
\restylefloat{figure}
\title{Spin correlation functions and 
susceptibilities in the easy-plane XXZ chain}
\author{C. Schindelin, H.~Fehske, H.~B\"uttner}
\address{Physikalisches Institut, Universit\"at Bayreuth, 
  D-95440 Bayreuth, Germany}
\author{D.~Ihle}
\address{Institut f\"ur Theoretische Physik, Universit\"at Leipzig,
  D-04109 Leipzig, Germany \\{\rm (\today)}
  }
\address{~\parbox{14cm}{\rm
    \medskip
We present a Green's-function theory of magnetic short-range order
in the $S=1/2$ easy-plane XXZ chain based on the projection method for 
the dynamic spin susceptibility  and a decoupling of three-spin
operator products introducing vertex parameters.   
The longitudinal and transverse static susceptibilities and two-point
correlation functions of arbitrary range are calculated self-consistently
for all  wavenumbers, temperatures, and anisotropy parameters 
$ -1\leq \Delta \leq 1$. 
In the easy-plane ferromagnetic region $(\Delta < 0)$,  
the longitudinal correlators of spins at distance $n$ 
change sign at a finite temperature $T_0(n,\Delta)$, 
in reasonable agreement with recent data 
obtained by finite-chain diagonalizations. 
The temperature dependence of the uniform static
susceptibilities exhibits a maximum which is explained as an effect
of magnetic short-range order which decreases with increasing temperature.
 }}
\begin{document}
\maketitle
The study of low--dimensional spin systems constitutes a growing field
of research which is mainly due to the availability of new materials,
such as the quasi--one--dimensional (1D) cuprates~\cite{Jo97}. 
Recently, an unexpected quantum--classical crossover of the longitudinal
spin correlation functions in the 1D XXZ model
\begin{equation}
  {\cal H} = \frac{J}{2}
\sum_{\langle i,j\rangle} (S_i^+S_j^- 
+\Delta  S_i^z S_j^z) 
\label{xxz}
\end{equation} 
($\langle i,j\rangle$ denote nearest--neighbor (NN) sites along the chain; 
throughout we set $J=1$) was found for  $-1<\Delta<0$ 
by means of exact diagonalization (ED) of systems up to 
18 spins~\cite{FM99} and by the quantum transfer matrix formalism~\cite{FKM99}.

Motivated by those findings, in this paper we examine the spin 
correlation functions in the easy--plane region  $-1<\Delta<1$ of the 
model~(\ref{xxz}) by an analytical approach
based on a Green's--function projection method. This theory was found
to provide a good description of antiferromagnetic (AFM) short--range order
(SRO) in the 2D spatially isotropic~\cite{WI97} and anisotropic Heisenberg 
models~\cite{ISWF99}. Moreover, for the first time, we 
calculate the full wavenumber, temperature and $\Delta$ dependences
of the static transverse and longitudinal spin susceptibilities.

 To determine the dynamic spin susceptibilities 
$\chi^{+-}(q,\omega)=-\langle\langle S_q^+;S_{-q}^-\rangle\rangle_{\omega}$
and  
$\chi^{zz}(q,\omega)=-\langle\langle S_q^z;S_{-q}^z\rangle\rangle_{\omega}$
$(-\pi\leq q\leq \pi)$, defined in terms of two--time retarded
commutator Green's functions, by the projection method, we choose the
two-operator basis  ${\bf A}_1=(S_q^+, i \dot{S}_q^+)^T$ and 
${\bf A}_2=(S_q^z, i \dot{S}_q^z)^T$, respectively, and consider 
the matrix Green's function 
$\langle\langle {\bf A};{\bf A}^{\dagger}\rangle\rangle_{\omega}$,
neglecting the self--energy part~\cite{WI97},
$\langle\langle {\bf A};{\bf A}^{\dagger}\rangle\rangle_{\omega}=
[\omega-{\frak M}'{\frak M}^{-1}]^{-1}{\frak M}$ with 
${\frak M}=\langle [{\bf A},{\bf A}^{\dagger}]\rangle$
and ${\frak M}'=\langle [i\dot{\bf A},{\bf A}^{\dagger}]\rangle$. 
We get
\begin{equation}
\chi^{\nu}(q,\omega)=-\frac{M^{\nu (1)}_{q}}{\omega^2
-(\omega_{q}^{\nu})^2}\,;\;\;\nu=+-,\,zz,
\label{chi}
\end{equation}
where the first spectral moments  
$M_{q}^{\nu (1)}$ are given by the exact expressions 
\begin{eqnarray}\label{mq1}
M_{q}^{+- (1)}&=&-2 [C_1^{+-}(1-\Delta\cos q)+2C_1^{zz} (\Delta-\cos q)]\,,\\
M_{q}^{zz (1)}&=&-2 C_1^{+-}(1-\cos q)\,. 
\label{mq2}
\end{eqnarray}
The two--spin correlation functions $C_n^{+-}=\langle S_0^+ S_n^- \rangle$
and $C_n^{zz}=\langle S_0^z S_n^z \rangle$ are calculated from
\begin{equation}
\label{cn}
C_n^{\nu}=\frac{1}{N}\sum_{q}\frac{M_{q}^{\nu (1)}}{2 \omega_{q}^{\nu}}
 [1+2p(\omega_{q}^{\nu})] \mbox{e}^{iqn}\,,
\end{equation}
where the Bose function 
$p(\omega_{q}^{\nu})=(\mbox{e}^{\omega_{q}^{\nu}/T}-1)^{-1}$
appears due to the use of commutator Green's functions.
The NN correlation functions are directly related to 
the internal energy per site $\varepsilon =C_1^{+-}+\Delta C_1^{zz}$.

To obtain the spectra $\omega_{q}^{\nu}$ in the approximations 
$-\ddot{S}^+_{q}=(\omega_{q}^{+-})^2 S^+_{q}$ and 
$-\ddot{S}^{z}_{q}=(\omega_{q}^{zz})^2 S^z_{q}$,
we take the site-representation and decouple the products 
of three spin operators in $-\ddot{S}^+_{i}$ and $-\ddot{S}^z_{i}$
along the NN sequence $\langle i,j,l\rangle$  
introducing vertex parameters in the spirit of the 
scheme proposed by Shimahara and Takada~\cite{ST91},
\begin{eqnarray}
S_i^+S_j^+S_l^-&=&\alpha^{+-}_1 \langle S_j^+S_l^- \rangle S_i^+
+\alpha_2^{+-}   \langle S_i^+S_l^- \rangle S_j^+\,,\\
S_i^zS_j^+S_l^-&=&\alpha^{zz}_1 \langle S_j^+S_l^- \rangle S_i^z\,,\\
S_i^+S_j^zS_l^-&=&\alpha^{zz}_2 \langle S_i^+S_l^- \rangle S_j^z\,.
\label{dcoup}
\end{eqnarray}
Here, $\alpha_1^{\nu}$ and $\alpha_2^{\nu}$ appearing in
$\omega_q^{\nu}$ are attached to correlation functions between 
nearest and further-distant neighbors  functions, respectively.
We obtain
\begin{eqnarray}
\label{w+-}
(\omega_{q}^{+-})^2&=&\frac{1}{2}[(1+2\alpha_2^{+-}C_2^{+-})
(1-\Delta \cos q)\nonumber\\[0.1cm]
&&\;\;+\Delta (1+4\alpha_2^{+-}C_2^{zz})(\Delta - \cos q)\nonumber\\[0.1cm]
&&\;\;+2\alpha_1^{+-}C_1^{+-}(\Delta \cos 2q - \cos q)\nonumber\\[0.1cm]
&&\;\;+4\alpha_1^{+-}C_1^{zz}(\cos 2q -\Delta \cos q)]\,,\\[0.2cm]
\label{wzz}
(\omega_{q}^{zz})^2&=&(1- \cos q)[1+2\alpha_2^{zz}C_2^{+-}\nonumber\\[0.1cm]
&&\;\;-2\Delta\alpha_1^{zz}C_1^{+-}(1 +2\cos q)]\,.
\end{eqnarray}
 
\begin{figure}[!htb]\caption{}
\epsfig{file= 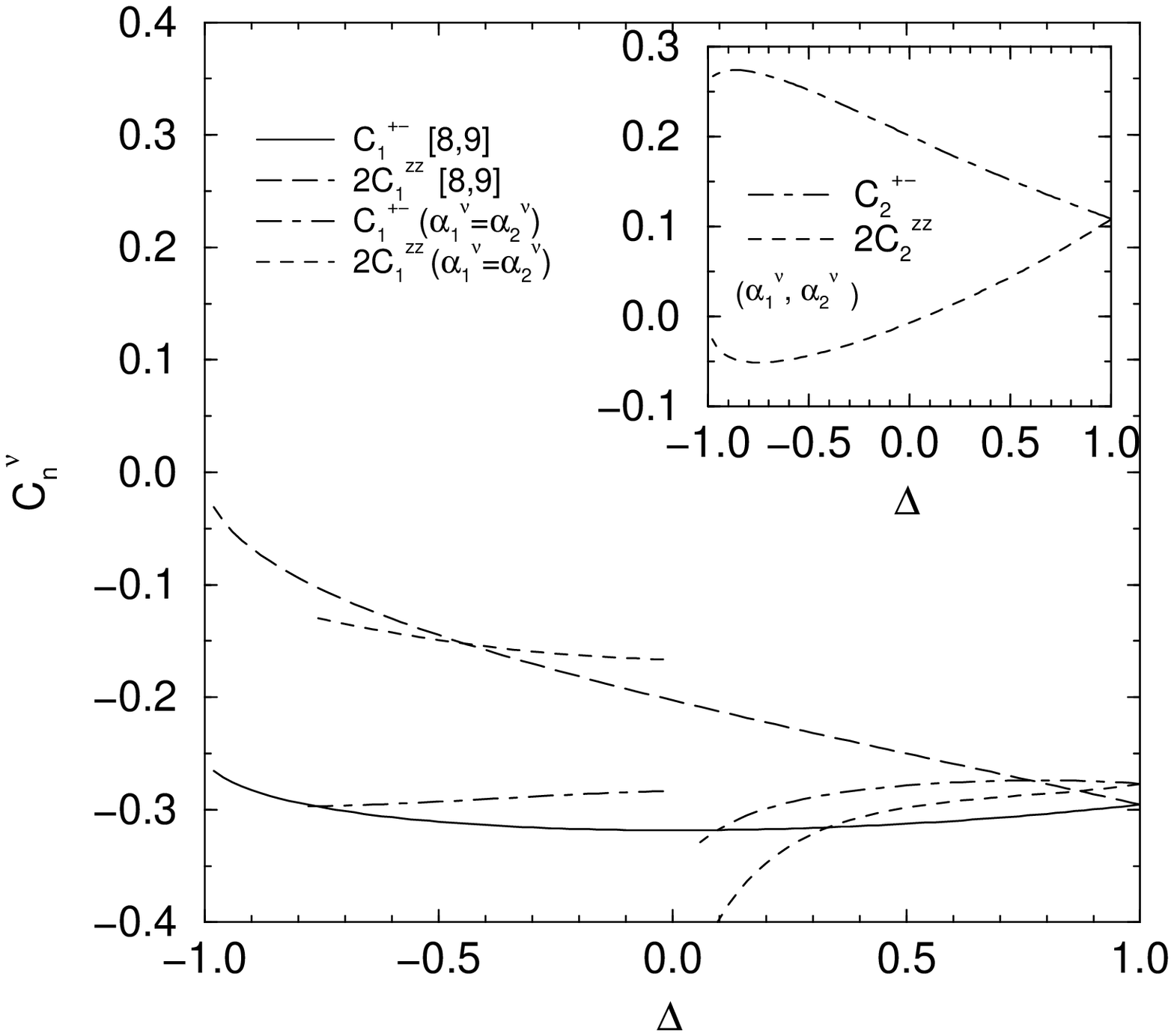, width = 1.0\linewidth}  
\label{fig1}
\end{figure}

In the easy-plane region $-1<\Delta<1$, the theory
has eight quantities to be determined self-consistently 
($C_1^{\nu}$, $C_2^{\nu}$, $\alpha_1^{\nu}$, $\alpha_2^{\nu}$) 
and six self-consistency equations~(\ref{cn})
including the sum rules $C_0^{+-}=1/2$ and $C_0^{zz}=1/4$. To get the two
remaining conditions for determining the free $\alpha$-parameters
$\alpha_2^{\nu}$ we may adopt different phenomenological
choices. Following the approach by Kondo and Yamajii~\cite{KY72}
for the isotropic Heisenberg chain ($\Delta =1$; 
$\alpha_i^{+-}=\alpha_i^{zz}$), let us first consider
the simple conditions $\alpha_2^{\nu}=\alpha_1^{\nu}$.
In this case we do not obtain quantitatively satisfactory results
(cf. Fig.~\ref{fig1}); therefore, we improve the theory by another choice 
with  $\alpha_2^{\nu}\neq \alpha_1^{\nu}$. For it we need additional 
conditions. At $T=0$, it is natural to 
adjust $\alpha_2^{\nu}$ to $C_1^{\nu}$ taken
from the exact expressions for the ground-state energy 
$\varepsilon (\Delta)$~\cite{YY66} and the NN correlator 
$C_1^{zz}$~\cite{JKM73} (see Fig.~1). Moreover,
to formulate conditions also at finite temperatures,
we follow the reasonings of Ref.~\cite{ST91} and~\cite{WI97}
and conjecture that both ``vertex corrections'' $ \alpha_1^{\nu}(T)-1$
and $ \alpha_2^{\nu}(T)-1$ have similar temperature dependences and
vanish in the high-temperature limit. Correspondingly, as the simplest
interpolation between high temperatures and $T=0$ we assume the ratio $R$
of two vertex corrections as temperature independent and given by the 
ground-state value. To be specific, in calculating  
$\chi^{zz}(q,\omega;T)$ for all $\Delta$, we assume 
\begin{equation}
\label{ratio1}
\frac{\alpha_2^{zz}(T)-1}{\alpha_1^{zz}(T)-1}
=R^{zz}\,.
\end{equation}
For $\chi^{+-}(q,\omega;T)$ the additional condition analogous to
Eq.~(\ref{ratio1}) (substitution of $\alpha_i^{zz}$ by $\alpha_i^{+-}$)
yields a solution of the self-consistency equations only in the case
 $\Delta > 0$. For  $\Delta < 0$ it turns out that a solution can 
be obtained using a slightly  modified condition (substitution of 
$\alpha_2^{zz}$  in Eq.~(\ref{ratio1}) by $\alpha_2^{+-}$). That is,
for $\chi^{+-}(q,\omega;T)$ we require
\begin{eqnarray}
\label{ratioa}
\frac{\alpha_2^{+-}(T)-1}{\alpha_1^{+-}(T)-1}
&=&R_{>}^{+-}\,,\\
\frac{\alpha_2^{+-}(T)-1}{\alpha_1^{zz}(T)-1}
&=&R_{<}^{+-}\,.
\label{ratiob}
\end{eqnarray}
The set of self-consistency equations is solved numerically
by Broyden's method with relative error less than 10$^{-7}$. 

To discuss the rotational invariance of the theory at $\Delta =\pm1$, 
it is useful to perform the unitary transformation 
which rotates the spins on all odd sites $m=2l+1$ 
around the $z$-axis by the angle $\pi$~\cite{IS88}. 
Taking the unitary operator
$U=\prod_{m}2 S_m^z$ which transforms ${\bf S}_n$ 
as $\tilde{\bf S}_n=U^+{\bf S}_nU$, we get
\begin{equation}
\label{utrafos}
 \tilde{S}_n^{x,y}=\mbox{e}^{i\pi n}S_n^{x,y}\,,\; \tilde{S}_n^{z}=S_n^{z}\,,\\
\end{equation}
and 
\begin{equation}
\label{utrafoh}
\tilde{H}=\frac{1}{2}\sum_{\langle i,j\rangle}(-S_i^+S_j^- +\Delta
S_i^zS_j^z)\,.
\end{equation} 
Since $\langle A\rangle_{\cal H}=\langle \tilde{A}\rangle_{\tilde{\cal H}}$
for any operator $A$, by~(\ref{utrafos}) we obtain the relation
\begin{equation}
\label{utrafox}
\chi^{+-}_{\cal H}(q,\omega)=\chi_{\tilde{\cal H}}^{+-}(k,\omega)\,;\;\;
k=q-\pi\,,
\end{equation} 
and $\chi^{zz}_{\cal H}(q,\omega)=\chi_{\tilde{\cal H}}^{zz}(q,\omega)$.
For the correlation functions we have $C^{+-}_{n,\cal H}=\mbox{e}^{i\pi n}
C^{+-}_{n,\tilde{\cal H}}$ and $C^{zz}_{n,\cal H}=C^{zz}_{n,\tilde{\cal H}}$.
Evidently, for $\Delta =1$ the rotational symmetry is preserved, i.e.,
$C_n^{+-}=2C_n^{zz}$, $\alpha_i^{+-}=\alpha_i^{zz}$, 
$\omega_q^{+-}=\omega_q^{zz}$, and 
$\chi^{+-}(q,\omega)=2\chi^{zz}(q,\omega)$.
At  $\Delta =-1$, the rotational symmetry occurs in the transformed
model~(\ref{utrafoh}), where the relations 
$\chi_{\tilde{\cal H}}^{+-}(q,\omega)=\chi^{+-}_{\cal H}(q+\pi,\omega)=
2\chi_{\tilde{\cal H}}^{zz}(q,\omega)$ and 
$2C^{zz}_{n,\tilde{\cal H}}=C^{+-}_{n,\tilde{\cal H}}$ 
can be easily verified from Eqs.~(\ref{mq1}), (\ref{mq2}), (\ref{w+-}), 
and~(\ref{wzz}). 

Corresponding to the unitary equivalence 
of $\tilde{\cal H}$ and ${\cal H}$ we denote
the (paramagnetic) easy-plane region with $-1<\Delta<0$ as ferromagnetic
(FM) region (cf. Eq.~(\ref{utrafoh})) and the region 
with $0<\Delta<1$ as AFM region (cf. Eq.~(\ref{xxz})).

Concerning the question of magnetic long-range order (LRO),
in the easy-plane XXZ chain there is no LRO which is correctly 
reproduced by our theory.  
In Ref.~\cite{ISWF99} the description of LRO
within our scheme (mode condensation) is outlined for the spatially 
anisotropic Heisenberg model, and the absence of LRO in the 1D limit 
is shown. Considering the model~(\ref{xxz})
at $|\Delta | <1$ and allowing for a possible finite
condensation part $C$ in the spin correlators~\cite{ISWF99}, 
a solution of our self-consistency equations only exists for $C=0$,
i.e., there is no LRO. 
 
In Fig.~1 the zero-temperature 
correlators appearing in the spectra~(\ref{w+-})
and~(\ref{wzz}) are plotted as functions of~$\Delta$. 
The NN correlators $C_1^{\nu}$, in particular
$C_1^{zz}$, calculated by the simplified theory 
($\alpha_2^{\nu}=\alpha_1^{\nu}$) strongly deviate from the exact
values~\cite{YY66,JKM73}, especially in the vicinity of $\Delta =0$.
The same refers to $\chi^{zz}(0)$. Therefore, in the following
we only show and discuss the results obtained by the improved theory
[cf. Eqs.~(\ref{ratio1}) to~(\ref{ratiob}).]    
At $\Delta =1$ the
rotational symmetry is visible. At the quantum critical point
$\Delta =-1$ there occurs FM LRO~\cite{IS88}, and 
a non-analytical behavior is observed, e.g.,  
$\lim_{\Delta \to -1^+} \partial C_1^{zz}/\partial \Delta=\infty$
and $\lim_{\Delta \to -1^-} \partial C_1^{zz}/\partial \Delta=0$~\cite{YY66}. 
Moreover, at $\Delta =-1$, we have $C^{+-}_{n,\tilde{\cal H}}=1/6$. 
This limiting behavior, however,  is hard to obtain 
numerically because of the infinite slope of
$C^{\nu}_{n}$ as $\Delta \to -1^+$. 

Figure~2 displays the static susceptibilities 
$\chi^{\nu}(q=0,\pi)$ at $T=0$ vs. $\Delta$.
First let us emphasize the excellent agreement of our result for the 
longitudinal uniform susceptibility with the exact 
result~\cite{Tak99} over the whole $\Delta$ region (cf. inset), 
where $\chi^{zz}(0)$ varies by two orders of
magnitude. That means, the longitudinal spin correlations of
arbitrary range are well described by our theory. 
In the FM region, the uniform longitudinal
susceptibility diverges in the limit $\Delta\to -1$ 
(cf. Fig.~5~(a)) indicating the instability of
the paramagnetic phase against the FM LRO phase at $\Delta =-1$~\cite{cond}. 
The same is true for the staggered transverse susceptibility
$\chi^{+-}_{\cal H}(\pi)$ (cf.  Fig.~5~(b)) which, by Eq.~(\ref{utrafox}),
is equivalent to the uniform transverse susceptibility 
$\chi^{+-}_{\tilde{\cal H}}(0)$.

The wavenumber dependences of the zero-temperature static susceptibilities
$\chi^{\nu}(q)$ are depicted in Fig.~3. For sufficiently low values 
of $\Delta$ in the FM region, $\chi^{zz}(q)$ shows a maximum at $q=0$ being 
indicative of the FM instability (cf.  Fig.~2). 
Accordingly $\chi^{+-}_{\cal H}(q)$  has a  maximum 
at $q=\pi$ (Fig.~3~(b)) corresponding, by Eq.~(\ref{utrafox}), to 
the maximum of $\chi^{+-}_{\tilde{\cal H}}(k)$
at $k=0$. In the AFM region, the longitudinal and transverse susceptibilities
reveal peaks at the AFM wavenumber being indicative of the 
AFM LRO at $\Delta >1$.

Let us now discuss the finite-temperature behavior of the longitudinal
spin correlation functions. Table~I shows that in the AFM 
region we get the expected alternating signs of $C_n^{zz}$ 
being characteristic of AFM SRO. By contrast, in the FM region,
$C_n^{zz}<0$ holds $\forall n$ below some characteristic  
temperature $T_0(\Delta)$~\cite{FM99}. In view of the approximations
made in our theory, the analytical results obtained 
for $C_n^{zz}$ are in reasonable agreement with the exact 
data for the 18-site chain. At fixed separation $n$ and with 
increasing temperature or above $T_0(\Delta)$ with increasing separation, 
$C_n^{zz}$ changes sign from negative to positive values. This property
can be interpreted as a quantum-classical crossover~\cite{FM99} because 
with increasing temperature the system behaves more classically,
i.e., it becomes dominated by the potential energy (longitudinal part 
of the Hamiltonian). It is worth emphasizing that the so-called 
``sign changing effect'' in the longitudinal spin correlations 
found numerically~\cite{FM99} is reproduced for the first time 
by an analytical theory. This is demonstrated in Table~II. 
The temperatures $T_0(n,\Delta)$ where $C_n^{zz}(T_0(n,\Delta),\Delta)=0$
are compared with the ED results showing an increasing 
agreement with decreasing anisotropy parameter $\Delta$. 
As might be expected,  for the 2D XXZ model our results
for $T_0({\bf r},\Delta)$ much better agree with 
the exact data (cf. Table~II). A detailed study of the 2D case 
is presented in Ref.~\cite{FSWBI00}.

Figure~4 shows the temperature dependence of the transverse correlation 
functions in the FM region. The oscillating 
behavior of $C_n^{+-}(T)$ (sign $(-1)^n$) and the decrease of $|C_n^{+-}|$
with increasing separation~$n$ (at fixed temperature) and increasing
temperature (at fixed $n$) indicates a transverse AFM SRO (but {\it no} LRO) 
related to the maximum of $\chi_{\cal H}^{+-}(q)$ at $q=\pi$ (cf. Fig.~3~(b)). 
However, due to the equality 
$C_{n,\cal H}^{+-}=(-1)^nC_{n,\tilde{\cal H}}^{+-}$,
for $\Delta <0$ we prefer to denote this SRO more physically 
as transverse FM SRO (in the model $\tilde{\cal H}$), where 
$C_{n,\tilde{\cal H}}^{+-}(T)$ can be read off immediately from Fig.~4.
Accordingly, for $\Delta >0$, we get transverse AFM SRO (similar behavior
of $C_{n}^{+-}(T)$ as shown in Fig.~4).

In Fig.~5 various susceptibilities $\chi^{\nu}(q)$ at $q=0,\,\pi$ are
plotted as functions of temperature. At $T=0$, the divergences of
$\chi_{\cal H}^{zz}(0)$ and 
$\chi_{\cal H}^{+-}(\pi)=\chi_{\tilde{\cal H}}^{+-}(0)$
in the limit $\Delta \to -1$, discussed above (Fig.~2),
are clearly seen. In the region $-0.3<\Delta<0$ the uniform longitudinal
susceptibility reveals a maximum at $T_{max}$ which may be  explained
as follows. For $T<T_0(\Delta,n)$ 
the longitudinal SRO, characterized by $C_n^{zz}< 0$
(cf. Table~I), leads to a spin stiffness against the orientation of the 
spins along a homogeneous external field in $z$-direction so that 
 $\chi^{zz}(0)$ is suppressed. With increasing temperature the correlations   
become increasingly ferromagnetic,  
which results in an increase of
$\chi^{zz}(0;T)$ up to $T_{max}$ determined by $T_0(n,\Delta)$. 
With decreasing $\Delta$ $(-1<\Delta < -0.3)$ the maximum disappears, since
the spin correlations are predominantly ferromagnetic. Thus, the maximum
in $\chi^{zz}(0;T)$ for  $-0.3<\Delta < 0$ may be understood as a combined
SRO and sign changing effect. In the AFM region, the maximum in the 
longitudinal susceptibility (see Fig.~5~(a)) obtained at all $\Delta$,
where $T_{max}$ shifts to higher values with increasing $\Delta$, is due
to the decrease of AFM SRO with increasing temperature.
At large enough values of $\Delta$, besides the maximum in $\chi^{zz}(0;T)$ 
there appears a minimum at a finite temperature which was also found 
for the isotropic Heisenberg chain~\cite{KY72} and contradicts the 
exact behavior~\cite{BF64}. Note that this artifact does not occur in the
2D XXZ model~\cite{FSWBI00}. In the high-temperature limit,
all susceptibilities depicted in Fig.~5 reveal a crossover to the 
Curie-Weiss behavior.

Finally, in Fig.~6 the temperature dependence of the susceptibility
$\chi_{\cal H}^{+-}(0)=\chi_{\tilde{\cal H}}^{+-}(\pi)$ is shown
which again may be explained as SRO effect. Here, in the FM region,
the transverse FM SRO results in a spin stiffness against the orientation 
of the transverse spin components along a staggered external field
perpendicular to the $z$-direction so that $\chi_{\tilde{\cal H}}^{+-}(\pi)$
is suppressed at low temperatures and exhibits a maximum.
In the AFM region, the transverse AFM SRO  results in
an analogous temperature dependence of $\chi_{\cal H}^{+-}(0;T)$,
where in the whole easy-plane region $T_{max}$ increases with increasing
$\Delta$.

To summarize, we presented a Green's-function theory of magnetic
SRO in the 1D easy-plane XXZ model which allows, for the first time,
the calculation of all static magnetic properties over the whole 
easy-plane FM and AFM regions, where no LRO occurs. 
That is, we computed the full wavenumber and 
temperature dependences of the anisotropic static spin susceptibilities 
and of the spin correlation functions of arbitrary range and 
at arbitrary temperature. In particular, in the FM region, 
we are able to reproduce the quantum-classical crossover 
in the longitudinal spin correlations in qualitative agreement 
with the numerical diagonalization data of Ref.~\cite{FM99}. 
Moreover, the obtained maxima in the temperature
dependences of the uniform longitudinal and transverse 
susceptibilities are explained as SRO effects. 
From the results of our theory we conclude that this approach 
may be successfully applied to other anisotropic spin models, 
where the description of spin correlations improves in higher 
dimensions~\cite{FSWBI00}.\\[0.2cm]

The authors would like to thank Y. Gaididei, 
J. Stolze, and A. Wei{\ss}e for helpful discussions.

\newpage
\section*{Figure Captions}
Fig. 1. Transverse and longitudinal spin correlation functions
$C_{n}^{\nu}$ ($n=1,2$) at $T=0$.\\[0.2cm]  

Fig. 2. Zero-temperature static susceptibilities at $q=0,\,\pi$.\\[0.2cm]

Fig. 3. Wavenumber dependence of the longitudinal~(a) and transverse~(b)
static susceptibilities at $T=0$.\\[0.2cm]

Fig. 4. Transverse correlation functions up to the fourth-nearest
neighbors in the easy-plane ferromagnetic region.\\[0.2cm]

Fig. 5. Inverse uniform longitudinal~(a) and staggered 
transverse susceptibility~(b).\\[0.2cm]

Fig. 6. Temperature dependence of the uniform transverse susceptibility.
\newpage
\onecolumn
\pagestyle{empty}
\begin{table}
\caption{Longitudinal spin correlation functions $C_n^{zz}(T;\Delta)$ 
for the easy-plane XXZ chain with $\Delta=\pm 0.3$ at $T=0.1$.
The ED data given in parenthesis are taken from Ref.~\protect\cite{FM99}.\\}
\begin{tabular}{ccccc}
$\;\Delta$ \rule{0mm}{4mm} &$n=1$ & $n=2$ & $n=3$ & $n=4$ \\[0.1cm] 
\hline
\rule{0mm}{4mm} $\;\;0.3$
& $-1.15\times 10^{-01}$ & $1.24\times 10^{-02}$ 
& $-9.60\times 10^{-03}$ & $1.34\times 10^{-04}$\\[0.15cm]
$-0.3$  & $-8.06\;[-8.38] \times 10^{-02}$ & $-1.31\;[-1.29]\times 10^{-02}$ 
& $-4.57\;[-8.08]\times 10^{-03}$ & $-1.89\;[-2.70] \times 10^{-03}$\\[0.1cm] 
\end{tabular}
\end{table}
\vspace*{2cm}
\begin{table}
\caption{Temperature $T_0(\Delta;{\bf r})$ of the sign change in the 
longitudinal correlation functions $C^{zz}_{\bf r}(T;\Delta)$ of the 
ferromagnetic easy-plane XXZ model. The corresponding results obtained
from ED of a 18-site 1D chain~\protect\cite{FM99} and a 2D 4$\times$4 square 
lattice with periodic boundary conditions are given in parenthesis.\\}
\begin{tabular}{cccccccc} 
  \rule{0mm}{4mm}$\Delta$ 
 & \multicolumn{7}{c}{$T_0(\Delta;{\bf r})$}  \\
 \rule{0mm}{4mm} & \multicolumn{4}{c}{1D case} &\multicolumn{3}{c}{2D case}\\
 \rule{0mm}{4mm} &  $n=1 $ & $n=2 $&  $n=3 $&  $ n=4 $& ${\bf r}=(1,0)$ 
& ${\bf r}=(1,1)$ & ${\bf r}=(2,0)$\\[0.1cm] \hline
  -0.1& 3.10 [4.966] &  2.18 [3.323]  & 1.72 [2.561] & 1.12 [2.073] 
& 2.98 [2.540] &1.76
  & 1.76 [1.520] \rule{0mm}{4mm} \\[0.1cm] 
 -0.3 &  0.86  [1.561] & 0.56 [1.071]& 0.44 [0.839] & 0.38 [0.687]
&0.96 [0.931] &  0.74 & 0.72 [0.713] \\[0.1cm] 
 -0.7 & 0.28 [0.413] & 0.20 [0.318] & 0.14 [0.264] & 0.12 [0.227]
&0.46 [0.391] &  0.36 [0.303]& 0.34 [0.301] \\[0.1cm]
 -0.9 & 0.12 [0.137] &    0.08 [0.118] & 0.06 [0.104] & 0.04 [0.092]
&$<$0.2 [0.125] & $<$0.2 [0.106]&$<$0.2 [0.106] 
\end{tabular}
\end{table}
\newpage
\newpage
\begin{figure}[!htb]\caption{}
\epsfig{file= 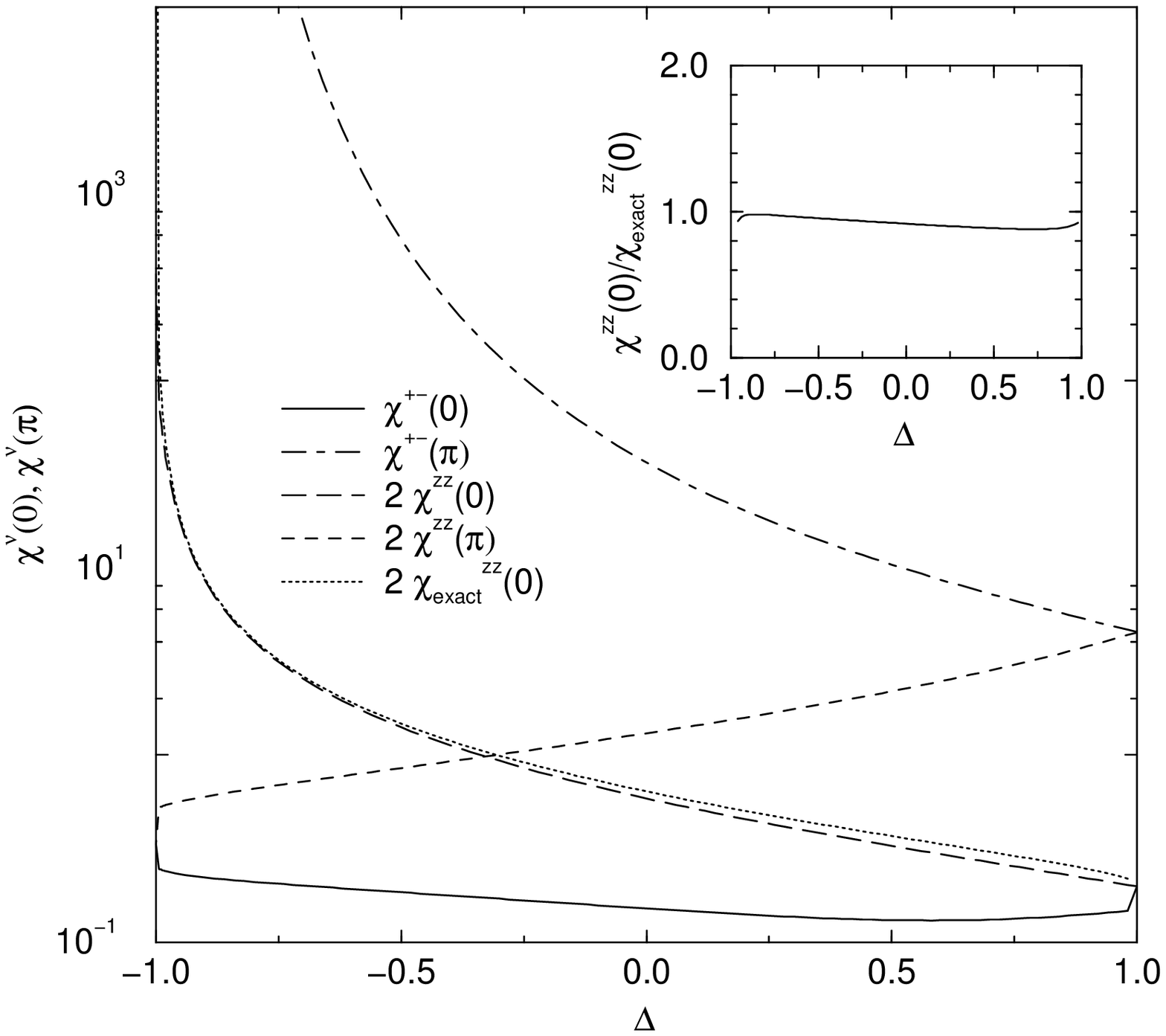, width = 1.0\linewidth}  
\label{fig2}
\end{figure}
\newpage
\begin{figure}[!htb]\caption{}
\epsfig{file= 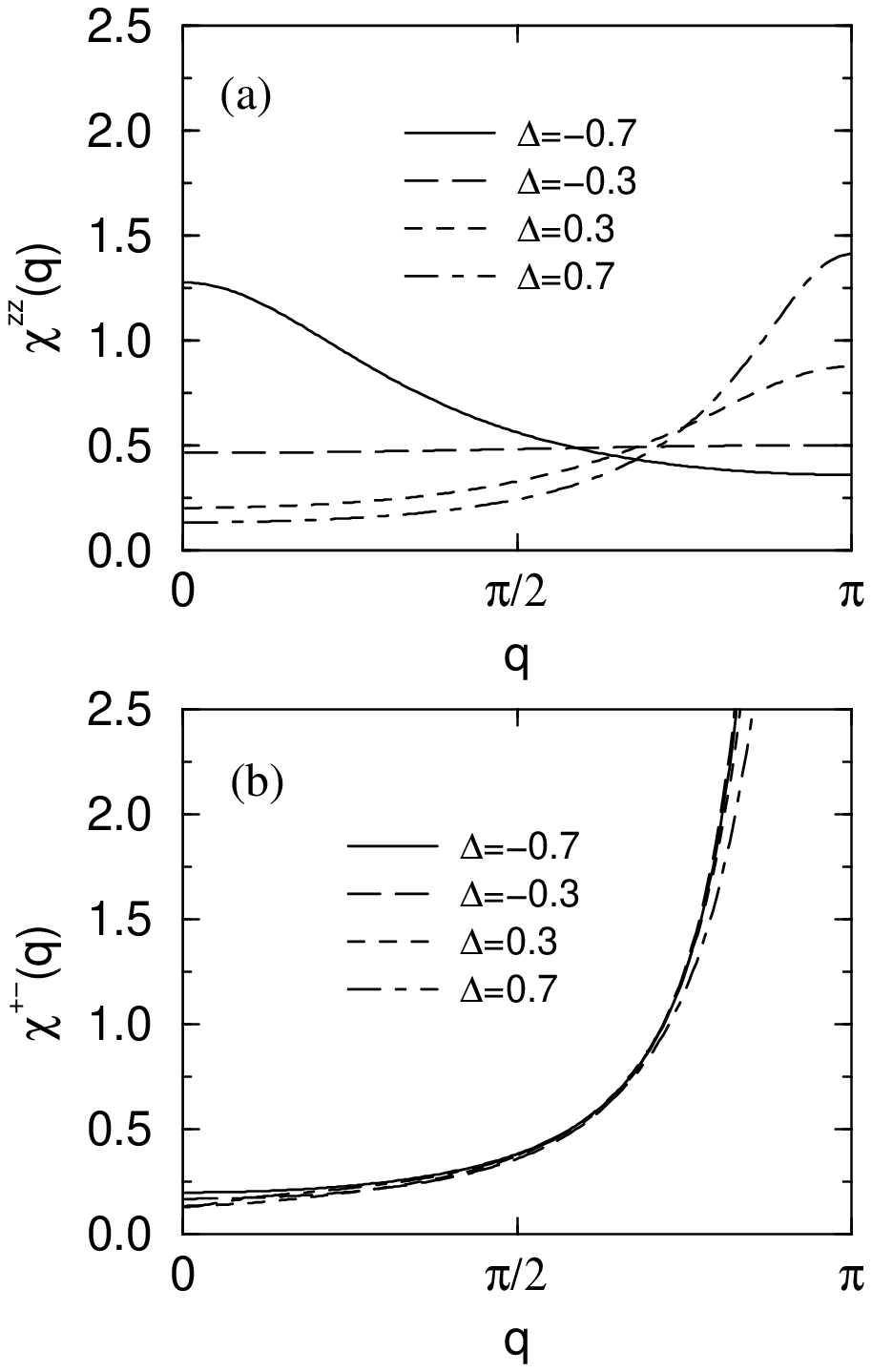, width = 0.8\linewidth}  
\label{fig3}
\end{figure}
\newpage
\begin{figure}[!htb]\caption{}
\epsfig{file= 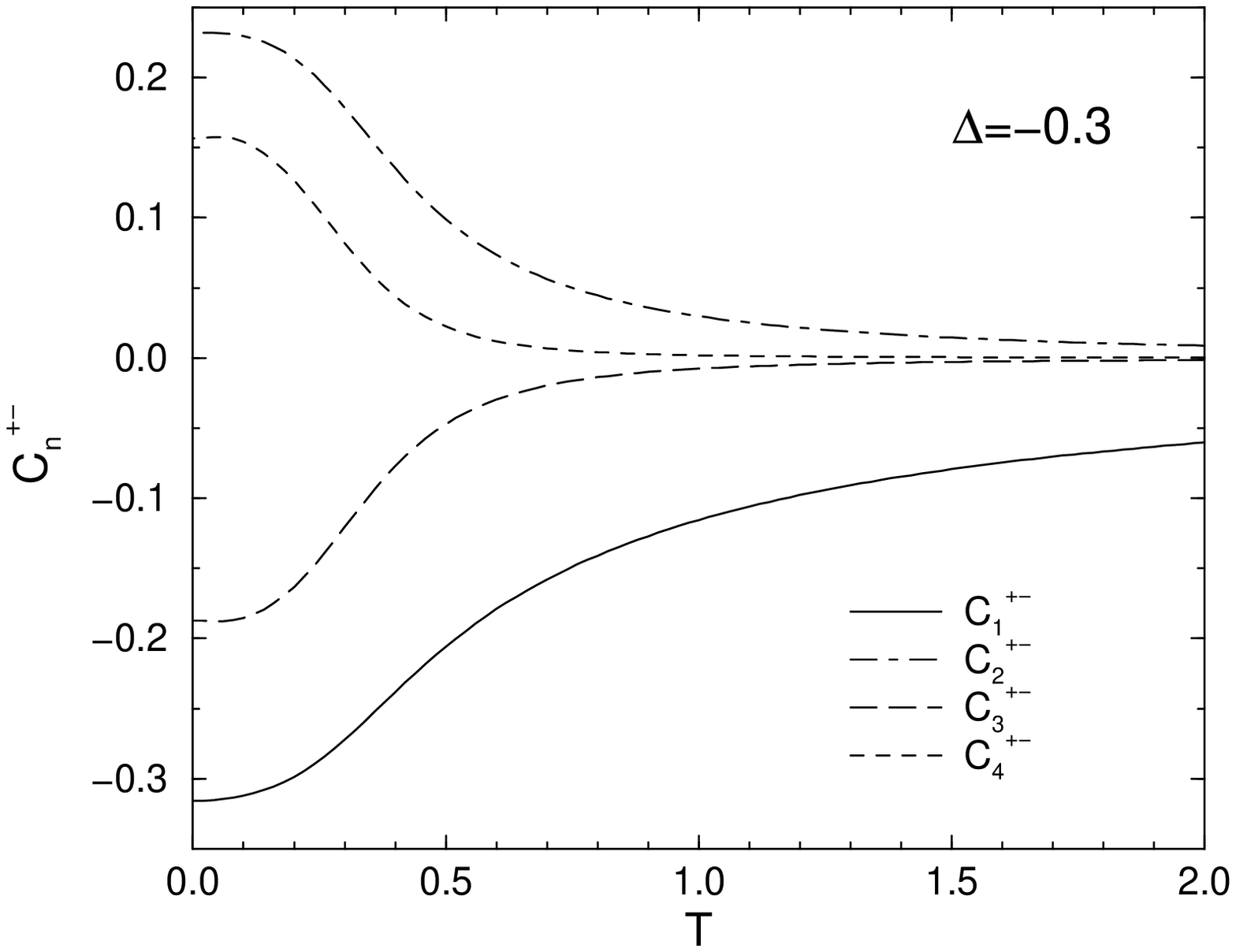, width = 0.9\linewidth}  
\label{fig5}
\end{figure}
\newpage
\begin{figure}[!htb]\caption{}
\epsfig{file= 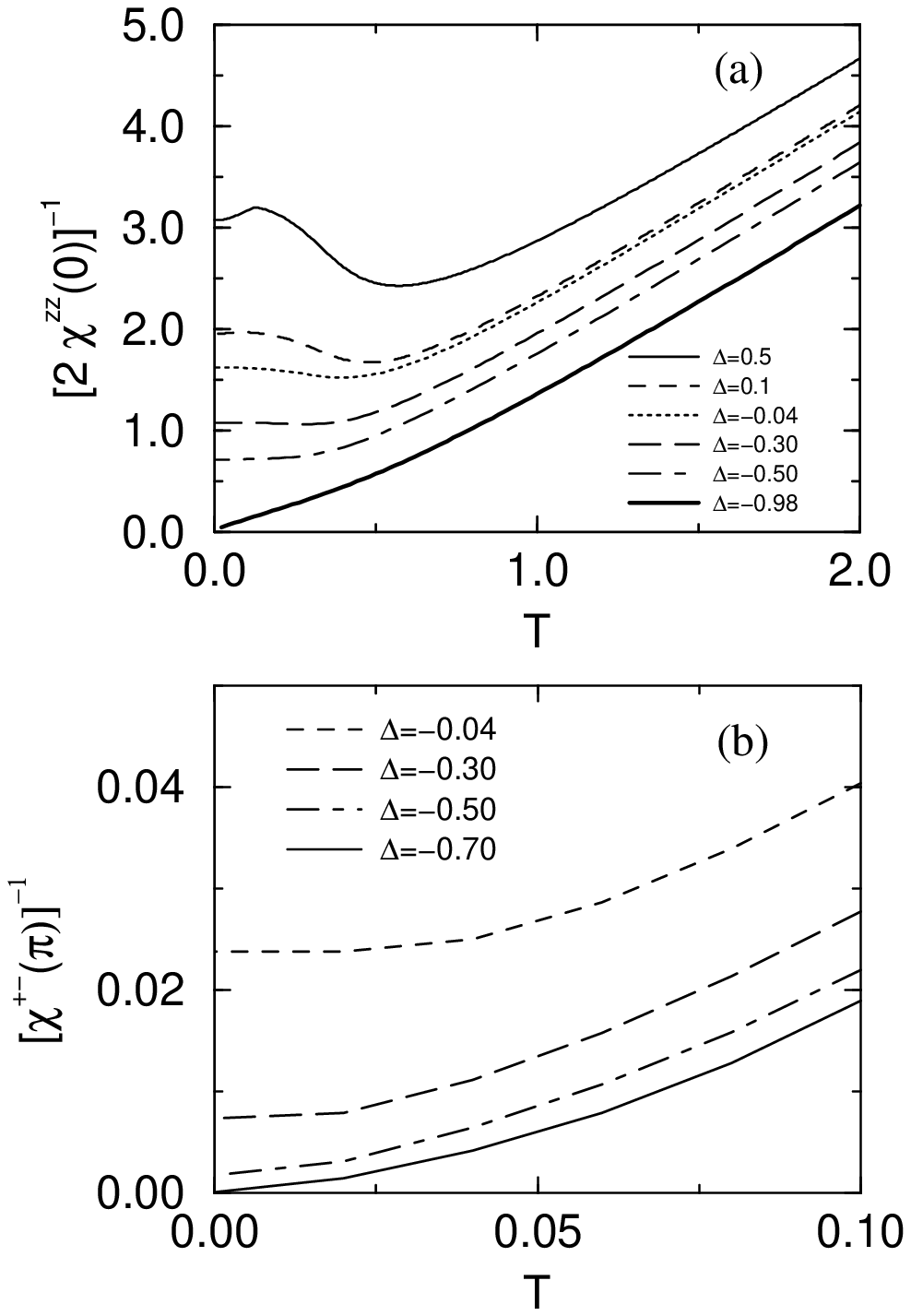, width = 0.9\linewidth}  
\label{fig6}
\end{figure}
\newpage
\begin{figure}[!htb]\caption{}
\epsfig{file= 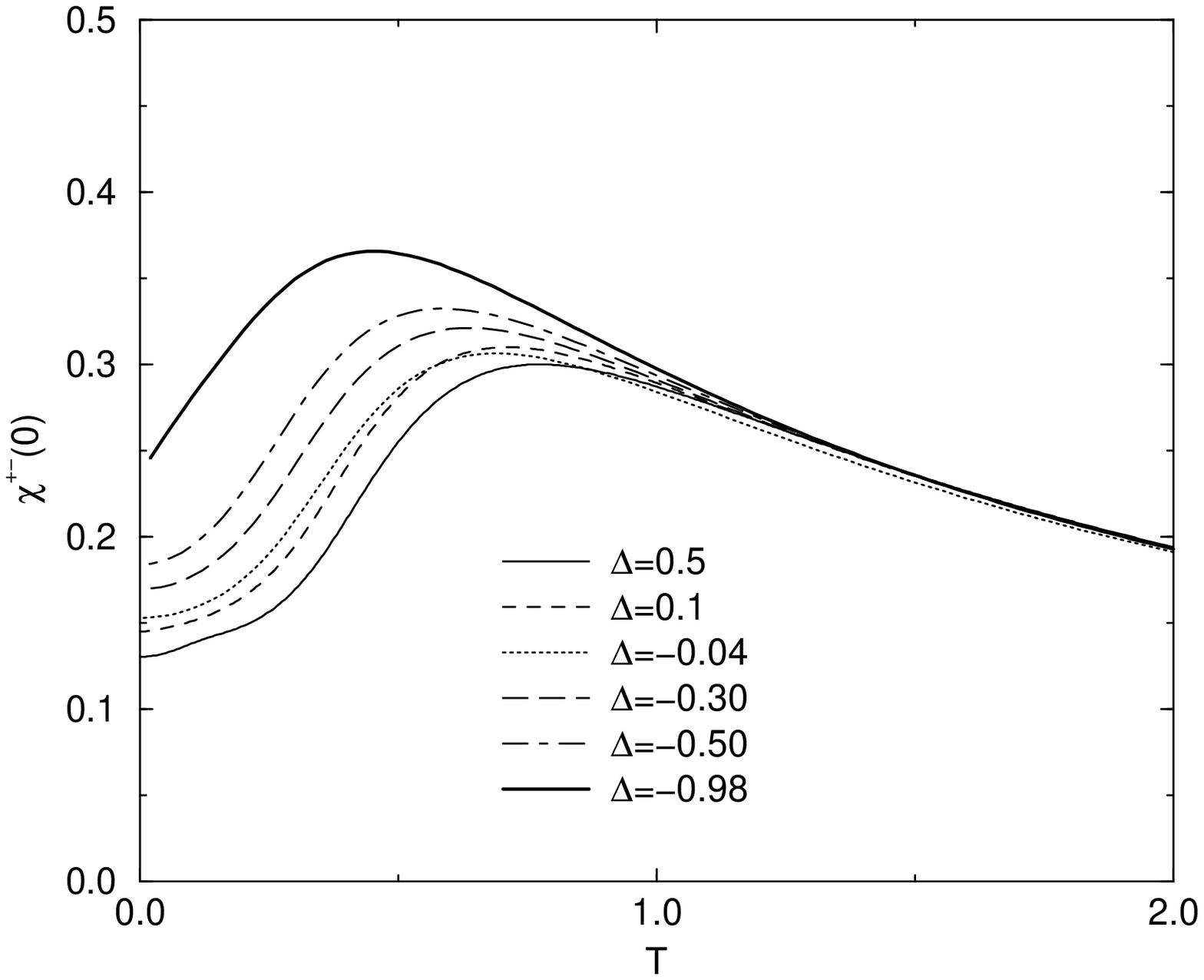, width = 1.0\linewidth}  
\label{fig7}
\end{figure}

\begin{thebibliography}{10}

\bibitem{Jo97}
D.~C. Johnston, in K.~H.~J. Buschow, editor, {\em Handbook of Magnetic
  Materials}, volume~10, Elsevier Science,  (Amsterdam 1997); 
A.~Tsvelik, {\em Quantum Field Theory in Condensed Matter}, Cambridge
  University Press,  (Cambridge 1995).

\bibitem{FM99}
K.~Fabricius and B.~M. McCoy, Phys. Rev. B {\bf 59}, 381 (1999).

\bibitem{FKM99}
K.~Fabricius, A.~Kl\"umper, and B.~M. McCoy, Phys. Rev. Lett. {\bf 82}, 5365
  (1999).

\bibitem{WI97}
S.~Winterfeldt and D.~Ihle, Phys. Rev. B {\bf 56}, 5535 (1997).

\bibitem{ISWF99}
D.~Ihle, C.~Schindelin, A.~Wei{\ss}e, and H.~Fehske, Phys. Rev. B {\bf 60},
  9240 (1999).

\bibitem{ST91}
H.~Shimahara and S.~Takada, J. Phys. Soc. Jpn. {\bf 60}, 2394 (1991);
{\it ibid.}  {\bf 61}, 989 (1992).


\bibitem{KY72}
J.~Kondo and K.~Yamaji, Prog. Theor. Phys. {\bf 47}, 807 (1972).


\bibitem{YY66}
C.~N. Yang and C.~P. Yang, Phys. Rev. {\bf 150}, 321 (1966);
{\it ibid.} 327 (1966);

\bibitem{JKM73}
J. D. Johnson, S. Krinsky, and B. M. McCoy, Phys. Rev. A {\bf 8}, 2526 (1973).

\bibitem{IS88}
Yu.~A. Izyumov and Yu.~N. Skryabin, {\em Statistical Mechanics of Magnetically
  Ordered Ssystems}, Consultants Bureau,  (New York 1988).

\bibitem{Tak99}
M. Takahashi, {\it Thermodynamics of one-dimensional solvable models},
Cambridge University Press (Cambridge 1999).

\bibitem{cond}
Note that for $\Delta \leq -1$ a condensation part will occur
within our scheme~\cite{ISWF99,ST91}.

\bibitem{FSWBI00}
H. Fehske, C.~Schindelin, A.~Wei{\ss}e, H. B\"uttner, and D.~Ihle,
cond-mat/0006272
\bibitem{BF64}
J. C. Bonner and M. E. Fisher, Phys. Rev. {\bf 135}, A640 (1964);
S. Eggert, I. Affleck, and M. Takahashi, Phys. Rev. Lett. {\bf 73},
332 (1994).

\end{thebibliography}
\end{document}